\documentclass[conference]{IEEEtran}
\usepackage[utf8]{inputenc}

\usepackage{csquotes}

\usepackage[numbers]{natbib}

\usepackage{graphicx}
\graphicspath{{images/}}

\usepackage{amsmath}
\usepackage{color}
\usepackage{tabularray}
\definecolor{CodGray}{rgb}{0.05,0.05,0.05}

\PassOptionsToPackage{hyphens}{url}\usepackage{hyperref}

\usepackage{array}
\renewcommand{\arraystretch}{1.35}

\usepackage{multirow}

\usepackage{float}

\usepackage{tcolorbox}

\usepackage{enumitem}

\usepackage{todonotes}

\def\BibTeX{{\rm B\kern-.05em{\sc i\kern-.025em b}\kern-.08em
    T\kern-.1667em\lower.7ex\hbox{E}\kern-.125emX}}

\begin{document}

\title{How Do Companies Manage the Environmental Sustainability of AI? An Interview Study About Green AI Efforts and Regulations}

\author{
    \IEEEauthorblockN{
        Ashmita Sampatsing\IEEEauthorrefmark{1},
        Sophie Vos\IEEEauthorrefmark{2},
        Emma Beauxis-Aussalet\IEEEauthorrefmark{1},
        Justus Bogner\IEEEauthorrefmark{1}
    }
    \IEEEauthorblockA{
        \IEEEauthorrefmark{1}Vrije Universiteit Amsterdam, Amsterdam, The Netherlands, e.m.a.l.beauxisaussalet@vu.nl, j.bogner@vu.nl
    }
    \IEEEauthorblockA{
        \IEEEauthorrefmark{2}Accenture, Amsterdam, The Netherlands, sophie.vos@accenture.com
    }
}

% \author{
%     \IEEEauthorblockN{Anonymous Author}
%     \IEEEauthorblockA{
%         Anonymous University\\
%         Anonymous City, Anonymous Country\\
%         anonymous@anonymous.com
%     }
%     \and
%    \IEEEauthorblockN{Anonymous Author}
%     \IEEEauthorblockA{
%         Anonymous University\\
%         Anonymous City, Anonymous Country\\
%         anonymous@anonymous.com
%     }
%     \and
%     \IEEEauthorblockN{Anonymous Author}
%     \IEEEauthorblockA{
%         Anonymous University\\
%         Anonymous City, Anonymous Country\\
%         anonymous@anonymous.com
%     }
% }

\maketitle

% \IEEEpeerreviewmaketitle

\begin{abstract}
With the ever-growing adoption of artificial intelligence (AI), AI-based software and its negative impact on the environment are no longer negligible.
Studying and mitigating this impact has become a critical area of research, as the adoption of AI technologies accelerates across various industries.
However, it is currently unclear which role environmental sustainability plays during AI adoption in industry and which Green AI practices users of AI-based software apply.
Moreover, little is known about how AI regulations influence Green AI practices and decision-making in industry.

We therefore aim to investigate the Green AI perception and management of industry practitioners.
To this end, we conducted a total of 11 interviews with participants from 10 different organizations that adopted AI-based software.
The interviews explored three main themes: AI adoption, current efforts in mitigating the negative environmental impact of AI, and the influence of the EU AI Act and the Corporate Sustainability Reporting Directive (CSRD) on the first two themes.

Our findings indicate that 9 of 11 participants prioritized business efficiency during AI adoption, with minimal consideration of environmental sustainability.
Monitoring and mitigation of AI's environmental impact were very limited.
Only a single participant monitored negative environmental effects, with four others at least tracking model token usage.
Regarding applied mitigation practices, six participants reported no actions, with the others sporadically mentioning techniques like prompt engineering, relying on smaller models, or not overusing AI.
Awareness and compliance with the EU AI Act are low, with only one participant reporting on its influence, while the CSRD drove sustainability reporting efforts primarily in larger companies, as noted by six participants.

All in all, our findings reflect a lack of urgency and priority for sustainable AI among these companies, with the main focus being the successful initial implementation and introduction of AI-based software.
We suggest that current regulations are not very effective in this regard, which has implications for policymakers.
Additionally, there is a need to raise industry awareness, but also to provide user-friendly techniques and tools to lower entry barriers for Green AI practices.
\end{abstract}

\begin{IEEEkeywords}
artificial intelligence, environmental sustainability, green AI, AI adoption, AI regulations, industry interviews
\end{IEEEkeywords}

\section{Introduction}
The capabilities and impact of artificial intelligence (AI)~\cite{flasinski_introduction_2016} have increased drastically over the last decade, leading to both excitement and worry worldwide.
A turning point in AI development occurred in November 2022 when OpenAI introduced a prototype of their chatbot ChatGPT~\cite{openai_chatgpt_2024}, a Generative Pre-trained Transformer (GPT) based on a Large Language Model (LLM).
Two months later, in January 2023, the Generative AI (GenAI) tool reached over 100 million monthly active users, making ChatGPT the fastest-growing consumer application in history~\cite{wu2023brief}.
However, training machine learning (ML) models like this and using them in complex applications at scale consumes vast amounts of energy~\cite{strubell2020energy}, thereby contributing substantially to carbon emissions.

The already achieved and further expected benefits of AI have led to an exponential increase in ML model size, bringing the environmental implications of AI to light~\cite{wu2022sustainable}.
For example, ChatGPT is built on GPT-4 foundation models containing over 1.76 trillion parameters, which is about 5,000 times as much as the Deep Neural Networks created in the early 2010s.
As a result, the associated energy consumption, and thus the carbon impact have increased dramatically~\cite{kirkpatrick2023carbon}.
\citet{chien2023reducing} estimated that ChatGPT-like chatbots produce 13 million kg of CO$_2$ per year, with an inference of 11 million prompts per hour.
Training these models requires over 50 million kWh, which can have a substantial negative impact on the environment~\cite{wu2022sustainable}.
As GenAI is becoming the fastest-growing technology in history, these emission rates are concerning.

Reducing and remediating climate change induced by green house gas emissions is critical and requires immediate and decisive action.
One of the key challenges in mitigating global warming is the transition towards more sustainable business, also known as the Green Transition~\cite{moodaley2023greenwashing}.
On the level of the United Nations (UN), the European Union (EU), or related regulatory bodies, this is often connected to Environmental Social Governance (ESG), i.e., the sustainability aspects that organizations are assessed on~\cite{eccles2020social}.
Over the last decade, there has been an increasing emphasis on Corporate Social Responsibility (CSR) among organizations, leading to increased interest in ESG reporting~\cite{aldogan2024determinants}.
For example, in January 2023, the Corporate Sustainability Reporting Directive (CSRD) entered into force\footnote{\url{https://eur-lex.europa.eu/eli/dir/2022/2464/oj}}, an EU regulation that establishes ESG reporting requirements for organizations.
This new legislation aims to improve the accessibility of ESG reporting, increasing trust and transparency in the sustainability reports of companies and organizations~\cite{birkmann2024csrd}.
Accordingly, sustainable development is expected to become a more important movement within companies globally.

Similarly, the EU AI Act\footnote{\url{https://eur-lex.europa.eu/eli/reg/2024/1689/oj}}, the first European legislation on AI to ensure the responsible development and usage of AI, entered into force in August 2024.
With the AI Act, the priority is to guarantee that AI-based software is safe, transparent, traceable, non-discriminatory, and environmentally friendly~\cite{fidon2024dempster}.
One of its key requirements is transparency and provision of information, i.e., companies have to become more transparent about how AI-based software is created and used, which may also help to assess its carbon emissions.
The AI Act requires record keeping, i.e., automatic logging during system operation, to ensure traceability throughout the application of AI-based software~\cite{wagner2023navigating}.
It also mentions environmental sustainability directly, albeit less forcefully.
For example, the AI Act refers to the principles of the EU's High-Level Expert Group for AI (one of them is societal and environmental well-being) and article 95 requires the EU AI Office to facilitate codes of conduct for voluntary application that include assessing and minimizing the environmental impact of AI.
All of this is meant to encourage companies to promote green and sustainable AI, aligning with the European Green Deal~\cite{dyrhauge2024introduction}.

One important step towards Green AI is to understand what companies are already doing to minimize the negative environmental impact of their AI-based software.
We currently have no clearly established overview of Green AI challenges and practices in industry, or of the reasons why specific actions are (not) taken.
Such insights are important to improve current practices and their adoption, but also to understand companies' Green AI motivations and barriers.
Moreover, such findings may also help to further understand which practices might (not) be applicable or effective for certain types of companies.
Lastly, it is important to look at these questions through the lens of the EU AI Act and the CSRD.
Understanding how these regulations influence companies will shed light on their effectiveness, but also on how aware companies are of their implications.
Currently, it is unclear how motivating or challenging practitioners perceive these regulations regarding Green AI initiatives in their work.
Therefore, we aim to provide insights into practitioners' perception of the environmental sustainability of AI and how this perception is impacted by both the EU AI Act and the CSRD.
More specifically, we answer the following research questions:

\textbf{RQ1:} How is a company's adoption of AI influenced by AI's negative impact on environmental sustainability?

For this RQ, we investigate whether choosing to adopt AI in a certain company has been influenced by the negative environmental impact of AI, and why this is (not) the case.

\textbf{RQ2:} How do companies adopting AI currently try to minimize its negative environmental impact?

For this RQ, we study a company's efforts to reduce the negative environmental impact of their AI-based software, e.g., which Green AI practices they apply and why.

\textbf{RQ3:} How do the EU AI act and the CSRD influence the adoption of AI and the mitigation of its environmental impact?

Finally, with RQ3, we investigate how these companies have changed or will change their stance on Green AI based on regulations, e.g., if they will apply new practices when both the AI Act and the CSRD have become fully active.

Our research bridges the gap between AI's negative environmental impact, its adoption by companies, and related regulations.
By examining mitigation measures and environmental considerations during AI adoption, we address an important knowledge gap and provide insights for designing and deploying Green AI practices or tools, as well as for policymaking.

\section{Background \& Related Work}
\label{sec:background}
As the significant energy demand for AI became evident, \citet{schwartz_green_2020} introduced the distinction between \textit{Red AI}, which prioritizes the highest accuracy without considering energy efficiency, and \textit{Green AI}, which aims to minimize its environmental impact while still achieving accurate results.
Recently, Green AI has emerged as an active research area, as highlighted in a literature review by \citet{verdecchia2023systematic} that identified 98 articles on the topic.
More than 75\% of these studies have been published since 2020.
The review categorizes the Green AI literature into 13 main topics, with the most prominent being monitoring, hyperparameter tuning, model benchmarking, deployment, and model comparison.
The authors suggest that interview studies could help understand how AI practitioners are currently addressing AI's environmental impact.
By synthesizing findings from our interviews with industry practitioners, we aim to provide a comprehensive understanding of the current landscape and to identify areas for future research in minimizing AI's environmental footprint.

\textbf{Research on the negative environmental impact of AI.} The energy consumption of machine learning models has been studied extensively in the last decade~\cite{sze2017hardware, yang2017designing, reagen2016minerva}.
For example, already in 2016, \citet{li2016evaluating} conducted a comprehensive evaluation of the energy efficiency of training frameworks for deep Convolutional Neural Networks (CNNs).
They presented a detailed study of the energy required for training using popular CNNs for image classification in computer vision, including a thorough analysis of different types of neural network layers.
Beyond energy, \citet{strubell2020energy} analyzed the carbon impact of training their own state-of-the-art models.
They concluded that developing and training such models incurs significant financial and environmental costs.
Financially, the costs come from the hardware and electricity required, and environmentally, they result from the carbon footprint required to fuel modern tensor processing hardware.

More recently, researchers examined the negative impact of Large Language Models (LLMs) on the environment, from the cost of training models to inference, highlighting the enormous recent and expected growth of AI.
\citet{wu2022sustainable} explained how the carbon footprint of LLMs is largely due to scaled-up inference, despite the substantial energy demands for their training.
As another example, \citet{luccioni2023estimating} discussed the carbon footprint of the BLOOM model, a 176-billion-parameter language model.
Using the Life Cycle Assessment (LCA) methodology, they evaluated the holistic environmental impact of the model.
Taking a broader perspective, \citet{luccioni2024power} provided a systematic comparison of the ongoing inference costs of AI models.
They found that the most energy- and carbon-intensive tasks are those that generate new content, such as text generation, text summarization, or image generation.
Some studies also estimated the carbon footprint of specific model architectures, such as GPT-3, Gopher, and OPT~\cite{patterson2021carbon, berthelot2024estimating, cursaru2024controlled}.
These studies focus primarily on the CO$_2$ emissions produced by LLMs.
However, the negative environmental impact of AI extends beyond CO$_2$ emissions.
For example, \citet{george2023environmental} investigated AI's water usage and showed that GPT-3 consumed 700,000 liters of water during its training phase alone.
The authors call for urgency to address this concern, as freshwater scarcity is a global problem~\cite{gleick2021freshwater}.

\textbf{Tools for monitoring the negative environmental impact of AI.} Another relevant research direction is the development of tools to measure or estimate the negative impacts of LLMs on the environment.
For example, \citet{budennyy2022eco2ai} introduced Eco2AI, an open-source package to track the energy consumption and CO$_2$ emissions of AI models.
Several similar open-source libraries are available to monitor CO$_2$ emissions during the training of AI models~\cite{henderson2020towards, dodge2022measuring, garcia2019estimation}.
Moreover, \citet{li2025makingaithirstyuncovering} presented formulas to estimate the water footprint of AI models, encouraging organizations to consider water usage when developing, training, or using AI.
Additionally, tools have been introduced to predict the carbon footprint of new neural networks before they have been trained. 
One early example is the Machine Learning Emissions Calculator by \citet{lacoste2019quantifying}.
A more recent calculator is LLMCarbon by \citet{faiz2023llmcarbon}, which is an end-to-end carbon footprint projection model designed for LLMs.

\textbf{Strategies for reducing the negative environmental impact of AI.} Many studies have been conducted to investigate strategies to reduce the energy consumption and carbon footprint of AI models and systems~\cite{verdecchia2023systematic}.
The research community has begun to translate this knowledge into reusable design decisions, such as architectural tactics~\cite{jarvenpaa_synthesis_2024}.
Another example is the study by \citet{yarally2023uncovering}, which examined the effect of Bayesian optimization during hyperparameter tuning on the energy consumption of model training.
They also investigated the complexity of AI models and concluded that the overall energy consumption of training can be halved by reducing the network complexity.
Hyperparameter tuning and model complexity have increasingly become topics of interest in recent years~\cite{panjapornpon2023explainable, kusumaraju24ecogen}.
Additionally, research has focused on improving the environmental sustainability of model inference.
For example, \citet{li2024toward} described how users can optimize their use of LLMs to minimize inference rates and reduce carbon emissions.

\textbf{AI adoption.} \citet{radhakrishnan2020determinants} conducted an in-depth analysis of 45 articles to identify the main theories and frameworks on AI adoption, including factors that facilitate, hinder, and determine the rate of diffusion in AI adoption in organizations.
They showed that the most dominant theories used to study AI adoption are Technology, Organization, and Environment (TOE) and Diffusion of Innovation (DOI).
At the organizational level, identified key barriers to AI adoption were digital maturity, trust, skill base, and privacy laws.
However, the study does not cover AI sustainability, e.g., its energy consumption or carbon footprint.

\textbf{Governance, laws, and regulations.} Recently, policymakers, practitioners, and academics have increasingly advocated for the development of global AI governance.
In a paper called \enquote{Digital Sovereignty, Digital Expansionism, and the Prospects for Global AI Governance}, \citet{roberts2023digital} argue that, while global governance initiatives to manage AI technology risks seem promising, they doubt the efficacy of such initiatives in practice.
They caution that aggressive pushes for digital sovereignty of individual countries may negatively impact such global governance initiatives, especially regarding the complex relationships between the EU, US, and China.
Moreover, \citet{birkmann2024csrd} studied the impact of the CSRD on large companies in Germany.
Through 14 interviews with communication experts of these companies, they explored how the CSRD has influenced CSR communication. They concluded that it remains uncertain whether the CSRD will lead to standardized CSR communications across different companies.

Despite extensive research on AI and its environmental footprint, there is still a significant gap in understanding the actions that companies are taking to mitigate the negative effects of AI on the environment.
This paper aims to start addressing this gap by examining how organizations using AI are working towards environmental sustainability in their AI practices and policies. 
Based on interviews with AI practitioners, we provide comprehensive insights into Green AI decision-making, applied practices, and areas for improvement.
Additionally, we examine the impact of regulations.
By focusing on real-world applications and actions, this study contributes to the existing literature by offering an analysis on how companies are striving to reduce the environmental footprint of AI technologies.

\section{Study Design}
Our research seeks to explore how companies perceive and manage the potential negative impact of their AI-based software on environmental sustainability.
Therefore, we wanted to study organizations already using AI-based software in their daily operations by conducting interviews with their employees.
To achieve this, we formed an academia-industry collaboration between our university and Accenture, a Fortune Global 500 professional services company specializing in IT and management consulting.
One of Accenture's key missions is about sustainable technology, i.e., supporting their clients in using the power of technology to drive sustainability transformations while continuously improving the sustainability of their used technology itself.
This partnership allowed us to ground our study design in industry needs, but also enabled us to make use of Accenture's partner and client network for participant recruitment.

More specifically, we relied on a qualitative approach using \textit{semi-structured interviews}~\cite{Hove2005}, allowing for an in-depth exploration of this topic.
Using semi-structured interviews provides rich and detailed results and also enables probing and discussing information that emerges from the participants' responses, leading to more in-depth information to be collected~\cite{bhattacherjee2012social}.
We used the Interview Protocol Refinement (IPR) framework by \citet{castillo2016preparing}, which consists of four phases to systematically develop an interview protocol:
1) ensuring interview questions align with research questions, 2) constructing an inquiry-based conversation, 3) receiving feedback on interview protocols, and 4) piloting the interview protocol.
To ensure transparency and replicability, we make our interview artifacts available on Zenodo.\footnote{\url{https://doi.org/10.5281/zenodo.14934325}}
This includes the interview preamble to recruit participants, the consent form, the interview guide itself, the slides used during the interview, and the key (anonymized) answers given by the participants.

\subsection{Sampling of Participants}
We imposed several criteria that participants had to meet to be included in the study.
First and most importantly, participants had to be employees of a company that actively uses AI-based software.
Additionally, participants should be able to provide information about the adoption and use of their AI-based software, and about the potential considerations of environmental sustainability in this context.
These criteria, while important for our RQs, led to a relatively small pool of potential participants, as employees may work with AI but are not always knowledgeable about how the company decides on the adoption of AI or what role sustainability plays~\cite{smith2024clinicians}.

To recruit participants, we mostly made use of \textit{convenience sampling}~\cite{Baltes2022}, i.e., we shared the call for participation within our personal networks via email and social media, as well as the network of Accenture.
Furthermore, the snowball technique was used to find more suitable interviewees for this research~\cite{biernacki1981snowball}, e.g., we asked participants to further distribute the call within their network.
Although such referrals are an effective method to recruit participants, the backgrounds of the referrers and their referrals can be quite similar, resulting in a sample with fairly homogeneous work experiences. 
Initially, we aimed to interview around 15 people.
However, over the course of the 5-month project, it became increasingly difficult to recruit more participants, mostly because of the requirement to have sufficient knowledge about the AI adoption and Green AI initiatives within the company or team.
Several practitioners were interested in participating but did not fully meet this criterion.
Consequently, our final sample consisted of 11 interviewees from 10 different companies, which still provides suitable diversity and breadth.

The participants included four women and seven men working at companies of varying sizes and domains in the Netherlands (Table~\ref{table:participant-demographics}).
Besides P10, all participants were native Dutch speakers.
Most of them had a background in IT, aligning well with the technology-focused nature of our research questions. Two participants had a background in both IT and Sustainability (P6, P7), offering interdisciplinary insights, and one participant had a unique professional background, namely in media, adding complementary perspectives (P5). 
The participants also had a wide variety of roles and hierarchical levels.
We recruited participants working in different industry domains with various levels of authority, seniority, information, and accountability regarding the adoption and usage of AI-base software.
Such diversity is highly valuable to gather different perspectives on AI decision processes and practices in industry.

\begin{table*}
    \centering
	\caption{Company and Participant Demographics (CID: Company ID, PID: Participant ID, YoE: Years of Experience)}
	\label{table:participant-demographics}
	\begin{tabular}{llrrp{5cm}rl}
		CID & Company Domain & \# of Employees & PID & Participant Role & YoE & Participant Background\\
		\hline
		\hline
		C1 & Banking & 10,000 - 50,000 & P1 & Green IT Consultant & 18 & IT\\
		\hline
        C2 & Data Science \& AI & 1 - 10 & P2 & Founder & 7 & IT\\
		\hline
        \multirow{3}{*}{C3} & \multirow{3}{*}{Consulting} & \multirow{3}{*}{$>$ 100,000} & P3 & \mbox{Managing$\,$Director$\,$Data$\,$\&$\,$AI} \mbox{within$\,$Strategy$\,$\&$\,$Consulting} & 18 & IT\\
		 & & & P6 & \mbox{Business Architecture} \mbox{Manager} & 10 & IT \& Sustainability\\
		\hline
        C4 & Semiconductors & 10,000 - 50,000 & P4 & Data and AI Analyst & $<$ 1 & IT\\
		\hline
        C5 & Humanitarian NGO & 101 - 1,000 & P5 & Product Owner & 12 & Media\\
		\hline
        C6 & Data Science \& AI & 1 - 10 & P7 & Co-Founder & 30 & IT \& Sustainability\\
		\hline
        C7 & Data Science \& AI & 1 - 10 & P8 & Founder & 3 & IT\\
		\hline
        C8 & Banking & 10,000 - 50,000 & P9 & Sustainable IT Lead & 25 & IT\\
		\hline
        C9 & Electronics \& Manufacturing & $>$ 100,000 & P10 & Software Architecture and Team Automation Lead & 7 & IT\\
		\hline
        C10 & Consulting & 10,000 - 50,000 & P11 & IT Consultant & $<$ 1 & IT\\
		\hline
	\end{tabular}
\end{table*}

\subsection{Study Execution \& Data Collection}
Before interviewing our 11 participants, one pilot interview was conducted to fine-tune the interview protocol, which is the fourth and final phase of the used IPR framework~\cite{castillo2016preparing}.
This pilot helped to improve the clarity of the phrasing and the order of the questions for better flow.
It also led to the addition of optional prompts to encourage more detailed responses.

According to the participants' availability, we conducted the interviews in April, May, and June 2024.
At the time of the interviews, the CSRD was already active (reporting for the calendar year 2024 needs to be provided in 2025), while the EU AI Act would enter into force slightly later in August 2024, with the first compliance requirements becoming active in early 2025.
All interviews were held in English.
As nearly all participants preferred an online interview over a face-to-face interview, we decided to do all interviews via Zoom.
All interviews were recorded with both sound and video.
They varied in length from 30 to 45 minutes.
During the interviews, the moderator made notes to write down any important follow-up questions and comments.

While the interviews were carried out in a conversational style, i.e., rather informal and open-ended, we still followed our interview guide.
During the interviews, it was important to ensure that the following terms were clear to all participants:
artificial intelligence, environmental sustainability, CSRD, and EU AI Act.
Therefore, we first presented a short slide set with the definitions of these terms to ensure a common understanding.
Afterward, we started with initial questions on the background of the participants and then moved to questions regarding the adoption of AI-based software in either their team or company.
To gain a broader perspective, we also asked about the primary drivers that usually lead to the decision of adopting AI.
We then explored how the negative environmental impact of AI affected its adoption.
Furthermore, the participant's current usage of AI was discussed, as well as the way environmental sustainability plays a role in this usage.
We were also specifically interested in Green AI practices that the companies have already applied or considered.
Finally, we checked how any of these topics were or will be influenced by the regulations of the CSRD and the EU AI Act.
For the complete interview guide, we refer to our replication package.

\subsection{Data Analysis}
After the interviews, we converted the recordings into transcripts.
For this, we used an AI-based transcription service called AIKO\footnote{\url{https://sindresorhus.com/aiko}}.
As this is an Apple-only application and all interviews were recorded on a Microsoft device, the recordings were first re-recorded by an Apple iPad, which was then used to transcribe them directly via AIKO.
However, to ensure accuracy of the transcripts, we had to review and refine them, as AIKO sometimes lacked accuracy with handling names, situations with noise in the background, or non-standard pronunciations of interviewees.
Additionally, the data collected in such a qualitative study includes more than words and textual information.
The participants' emotions, tone of voice, facial expressions, or attitudes can also provide important context.
We thus noted down the most important of such observations after reviewing the recordings.
The transcripts, together with these observations, were used for further analysis.

Once the transcripts were final, we applied thematic analysis~\cite{cruzes2011recommended}.
We first removed irrelevant information from the transcripts, such as words that had to be repeated due to mispronunciation or noise, or statements that did not answer any interview questions and were irrelevant for our RQs.
We then annotated the pre-processed data to create a codebook that structured our analysis.
A popular qualitative analysis tool, MAXQDA, was used to analyze the pre-processed data further~\cite{maxqda}.
We created labels that aligned with our RQs, e.g., the RQ1 analysis (consideration of sustainability during AI adoption) led to labels about drivers or identified concerns.
The codebook made it easier to merge labels that contained similar information and to create categories.
Overall, this was a lengthy iterative process, with labels being continuously refined, merged, and split.
As the last step, we elicited hierarchically ordered categories and characterized their relationships to each other.

\subsection{Ethical Considerations}
Before the interviews, we clearly explained to the participants what data will be collected and what we will do with it.
We assured them that we will keep the interview data confidential,  all published information will be aggregated or anonymous, and that specific participant privacy requirements regarding the data reporting will always be considered.
Finally, we emphasized that their participation was completely voluntary and that they could withdraw from the study at any point.
Before we began the interviews, the participants had to sign a consent form outlining these terms to ensure their protection and the integrity of the study.

\section{Results}
In this section, we present the findings from our interviews.
The analysis led to the creation of three themes, namely AI adoption (RQ1), environmental impact mitigation (RQ2), influence of regulations (RQ3).
These themes were further organized into eight subthemes.

\subsection{Sustainability Considerations During AI Adoption (RQ1)}
The first theme concerns the process of AI adoption within companies and what role environmental sustainability plays during the decision-making.
Table~\ref{table:ai-adoption-overview} provides an overview of all associated labels.
Self-employed participants or founders had full autonomy over AI adoption in their work (P2, P7, P8).
P5 was responsible for AI adoption together with the rest of their team.
The other participants could not completely describe all involved parties in the companies’ process of AI adoption (P1, P3, P4, P6, P9, P10, P11).
All participants referred to a decisional body, e.g., an AI council, for such decisions.
For instance, Sustainable IT Lead P9 said: \enquote{We have an AI board. So there are a lot of people, a lot of managers sitting together from different areas discussing what we can use from AI, where it makes sense, and also the legal consequences.}

\begin{table}[ht]
\renewcommand{\arraystretch}{1.2}
    \centering
	\caption{AI Adoption: Subthemes and Labels (RQ1)}
	\label{table:ai-adoption-overview}
	\begin{tabular}{lp{5.7cm}}
		Subtheme & Label\\
            \hline
            \hline
            \multirow{5}{*}{Primary Drivers} & Efficiency [P2, P3, P5, P6, P7, P8, P9, P10, P11] \\
                                         & Hype [P1, P4, P9, P10]                                  \\
                                         & Personal Interest [P1, P7]                              \\
                                         & Security [P9, P11]                                      \\
                                         & Quality [P5]                                            \\
            \hline        
            \multirow{11}{*}{AI Use Cases} & Chatbot [P1, P5, P6, P9] \\
                                         & Data Processing and Analysis [P2, P3, P10, P11]         \\
                                         & Code Writing [P2, P7, P8]                               \\
                                         & Automation [P4, P11]                                    \\
                                         & Recognition [P8]                                        \\
                                         & Recommendation [P6]                                     \\
                                         & Forecasting [P10]                                       \\
                                         & Building Custom GenAI Tooling [P7]                      \\
                                         & Translating [P5]                                        \\
                                         & Summarizing [P9]                                        \\
                                         & AI for Sustainability [P3]                              \\
            \hline
            \multirow{4}{*}{Identified Concerns} & Carbon Footprint [P2, P3, P4, P8, P10] \\
                                         & Energy Consumption [P1, P5, P7, P11] \\
                                         & Ignorance of True Impact [P6, P9]              \\
                                         & Water Use [P1]                                         \\
            \hline
	\end{tabular}
\end{table}

\textbf{Primary Drivers:}
Nine participants mentioned efficiency as the primary driver for adopting AI (P2, P3, P5, P6, P7, P8, P9, P10, P11).
The combination of saving time and costs was considered to lead to more efficient work processes.
For example, P6 said: \enquote{Well, in this case, it is an efficiency game. The large company I work for is operating globally. The biggest challenge we have is how to centralize the organisation in such a way that we are benefiting from the global network. Therefore, we have adopted AI in a reinvention console to reach out to people globally.}
Similarly, P7 said: \enquote{So, the reason why we use AI is because automation in terms of just development of normal IT tooling is either impossible or too expensive.}
Finally, P10 said: \enquote{The primary drivers are to make it faster that we do not need so many engineers to do the work.}
Other drivers for adopting AI were personal passion (P1, P7), higher quality (P5), better security (P9, P11), and the current hype around AI (P1, P4, P9, P10).
P9 mentioned the following on using AI for its hype: \enquote{Now, there is a huge demand for GenAI solutions because it is a technique that we can use to create natural language, like, for example, for chatbots. Since I joined in the past few months, we have only done GenAI things. Because it is now just really a complete hype trend, whatever you want to call it.}

\textbf{AI Use Cases:}
All participants used at least one off-the-shelf AI tool created by big tech companies such as Google, OpenAI, and Microsoft.
P4, P7, P8, and P10 also used internally created AI-based software.
The most frequently mentioned AI usage scenarios were for processing and analyzing data (P2, P3, P10, P11) and chatbots (P1, P5, P6, P9).
Furthermore, AI was used to write code (P2, P7, P8), to automate processes (P4, P11), to recognize patterns (P8), as a recommendation algorithm (P6), to forecast (P10), to build custom GenAI tooling (P7), and to translate (P5) and summarize text (P9).
Founder P2 primarily used AI to process data: \enquote{What we are looking into more and more is to let AI do our analysis for us. So, we provide some data to an AI model, and it creates some insights for our customers.}

Furthermore, Managing Director P3 mentioned a case in which AI was used to increase environmental sustainability: \enquote{A client of mine has an initiative to support smallholder farmers in Asia, Africa, and South America to help them with agroforestry to build a second earning model. Using satellites and AI, farms are kept an eye on to check the biomass growth and to see whether that is not too much or too little to feed people living close by. This is called carbon removal units.}

\textbf{Identified Concerns:}
All participants were aware of the negative environmental impact of the AI-based software they adopted or used.
When they referred to this impact, five participants focused on the carbon footprint of AI (P2, P3, P4, P8, P10), while four focused on its substantial energy use (P1, P5, P7, P11).
For example, Product Owner P5 said: \enquote{I know that just one prompt, just one question to an AI chatbot, has the same footprint as the use of one plastic bag.}
Founder P2 was concerned with the carbon emitted when developing and training AI: \enquote{I see AI as a form of data analysis by itself, and data analysis has a negative impact on the environment. We store a lot of data, and those databases must run somewhere. They need a lot of energy, and by combining it with AI, even more energy is being used to run this entire operation. The models are way larger. We already were ruining the environment, and now we are doing worse times 10.}
Managing Director P3 mentioned the use of AI for sustainability, but was also aware of the negative environmental impact: \enquote{It is like flying around the world to ensure we get a better climate. It is a bit double standard}.

However, some participants also were concerned that their organizations did not take the negative environmental impact of AI seriously enough or were ignorant of its true impact.
Business Architecture Manager P6 said: \enquote{However, I am also pretty convinced that of the money which is reserved, which we are using internally to adopt and use AI, a large portion of that is not taking into consideration the environmental footprint.}
Similarly, Sustainable IT Lead P9 said: \enquote{And the AI board said they had a look at the environmental impact of AI, and said it was very small, in their opinion. I disagree with that opinion.}
Finally, P1 highlighted the water use of AI but immediately tried to justify this point: \enquote{Of course, the more power is being used, the more water it needs. But on the other hand, it is way more efficient and faster.}

\subsection{Mitigating the Environmental Impact of AI (RQ2)}
When analyzing how companies currently try to minimize the negative environmental impact of AI, we identified three subthemes: monitoring AI's impact, mitigating efforts, and associated challenges.
Table~\ref{table:mitigation-overview} provides an overview of all created labels.

\begin{table}[ht]
\renewcommand{\arraystretch}{1.2}
    \centering
	\caption{Mitigation Efforts: Subthemes and Labels (RQ2)}
	\label{table:mitigation-overview}
	\begin{tabular}{lp{5.7cm}}
		Subtheme & Label\\
            \hline
            \hline
            \multirow{3}{*}{Monitoring} & None [P1, P2, P3, P4, P5, P6, P7, P8, P10, P11] \\
                & Monitoring Other Tokens [P1, P4, P7, P11] \\
                & Cloud Carbon [P9] \\
            \hline
            \multirow{7}{*}{Mitigating Efforts} & None [P1, P2, P5, P6, P8, P9] \\
                & Not Overusing AI [P3, P7, P10] \\
                & Not Using Unnecessary Large Models [P4, P11]                                                                                                               \\
                & Prompt Engineering [P4]                                                                                                                                    \\
                & Reusing AI Solutions [P7]                                                                                                                                  \\
                & Caching [P7]                                                                                                                                               \\
                & Considering Location of Running AI Model [P10]                                                                                                             \\
            \hline
            \multirow{6}{*}{Green AI Challenges} & No Transparency and/or Responsibility from Big Tech Companies [P2, P3, P4, P7, P8] \\
                & Lower Priority [P5, P9, P10] \\
                & No Budget [P1] \\
                & Gap between Sustainability Department and IT Department [P11]  \\
            \hline
	\end{tabular}
\end{table}

\textbf{Monitoring:}
Sustainable IT Lead P9 was the only participant who stated that his company monitored the negative environmental impact of their AI-based software in some form, namely its carbon footprint via the tool CloudCarbon.
The rest of the interviewees did not monitor (any part of) the negative environmental impact of their AI-based software, without providing a clear reason for not doing it.
For example, Managing Director P3 stated: \enquote{Everybody is aware of the carbon footprint of AI. But it has not yet resulted in monitoring.}
However, while P1, P4, P7, and P11 explained that they did not explicitly monitor the negative environmental impact of AI, they instead monitored the used LLM tokens, which eventually could help to estimate the carbon footprint.
For example, P7 explained: \enquote{No, we are not monitoring the negative environmental impact of the AI tools we use. The thing we are using, which is a proxy for it, is the number of tokens, better said, the number of computes.}
Similarly, P4 said: \enquote{We are currently working a bit more on monitoring, not specifically for the environment yet. Still, we are now logging and monitoring, for example, how many tokens are used as input and output from these models. We could also use that to monitor environments, maybe emissions. I think we have the information present of how many times we use the models and with which information it could be possible to calculate in the future. However, right now, it is not calculated.}
P11 mentioned that their organization monitored their AI tools to track and improve their cost efficiency: \enquote{The AI tools are monitored to essentially keep track of the cost efficiency of what we are running. But when you track it in Euros, you can also basically translate that to CO$_2$ costs. However, we do not do this yet.}

\textbf{Mitigation Efforts:}
Regarding companies' current efforts to reduce the environmental impact of AI, six participants mentioned that they currently make no mitigating efforts (P1, P2, P5, P6, P8, P9).
The other five participants reported different mitigation practices.
For example, prompt engineering was applied by the team of P4, caching was applied by P7, who also tried to reuse AI solutions, and P10 practiced picking the most environmental friendly location to run AI models.
Data and AI Analyst P4 explained the systematic use of prompt engineering: \enquote{So, we give prompt engineering workshops internally to make people aware that if they ask a stupid question, they will get a stupid answer, mostly. So then, of course, if they have to prompt many times, that will all be calls to the model with emissions.}
Furthermore, P4 and P11 avoided using unnecessary large models for tasks that could be carried out by smaller and less energy-hungry models.
IT Consultant P11 used this example: \enquote{If we know that questions can be answered with GPT 3.5 Turbo as well, which is a lighter model than GPT 4, then we can use GPT 3.5 Turbo.}
Lastly, the most mentioned practice was to not overuse AI (P3, P7, P10).
Co-Founder P7 said the following about this: \enquote{We see AI in general as a tool. Once you have a hammer, not everything is a nail. So, you should also apply it to those kinds of problems where it excels. So, I see a lot of people trying to apply AI chatbots to everything, also to things for which it is not meant, for which it does not excel, for which there are other solutions that are much better, much more efficient.}
Similarly, Managing Director P3 said: \enquote{If it is not necessary, then we do not use AI. Then we use a dashboard, for instance. And well, you can see this as a decreasing footprint.}
However, P10 finished her related answer with a question emphasizing that determining this necessity is not always straightforward: \enquote{And in general, how often should we use LLMs?}

Finally, when talking about monitoring and mitigation efforts, the words \enquote{not yet} were regularly used by participants (P3, P4, P5, P10, P11).
For example, P10 said: \enquote{I think we are just not there yet. We are at the beginning of putting AI into our products and so on, but we still need to understand how to do it. Yeah, and then I guess it is the next step to see how we can do it best}.
Likewise, P5 said: \enquote{No, so it is really premature, the use of AI. And I do not have the insights yet. Because the first challenge is to have a working technical product. So, we are still in the process of making it meet the user and business requirements and technical requirements, and after that has been done, we will focus on the sustainability of the tools.}

\textbf{Challenges:}
Most interviewees mentioned at least one challenge that arose while trying to move towards Green AI.
Four participants using off-the-shelf AI tools reported that big tech companies lack transparency regarding the environmental impact of the AI tools they offer (P2, P3, P4, P7, P8).
For example, Co-Founder P7 described it in the following way: \enquote{I think the biggest problem is with these large AI companies that we use these models from. I do not think they are very elaborate or very extensively documenting things that we can also do to measure our impact or maybe reduce it more. I also do not want to put all the blame on them because I think people using it should also take it into account. But I think that is definitely something for which there is not much attention.}
Similarly, P8 said: \enquote{What I find really hard in this kind of discussion is that it is hidden for you as well, right? You pay a certain price to Google and OpenAI, which use a lot of energy to create these AI models. And for their client to provide insights into our AI footprint, these big tech companies should make this more visible. Right now, you do not really see them, and you cannot make it really tangible how much CO$_2$ a specific question just generated.}
Lastly, Founder P2 concluded: \enquote{So, because we are not hosting our own hardware, I do not feel in control of actually doing anything with this. I have to rely on services like Microsoft to implement this in the right way.}

Furthermore, P5, P9, and P10 expressed how AI sustainability simply had lower priority than other concerns in their companies, or as P10 put it: \enquote{I would still say it is a secondary point at the moment.}
Likewise, Sustainable IT Lead P9 said: \enquote{So, I would say, so far, we have been more on the opportunity side. Like I said earlier, productivity gains or efficiency gains. And so the real guardrails around limiting the environmental impact of AI are still to be set, I should admit. And this is where our conversations are also about.}

Another challenge was reported by IT Consultant P11, who mentioned a gap between the IT department and other colleagues focusing on sustainability: \enquote{I have noticed that there is quite a large disconnect between, whatever the sustainability people are doing and whatever the IT people are doing. There is not much of a bridge there. The IT people have this nagging feeling inside somewhere, saying what I am doing might be harming the environment, and they want to do better, but they do not know what exactly. Whereas the sustainability people generally know what should be done to change this, but they do not know how to directly implement such things.}
Lastly, Green IT Consultant P1 explained that there was no extra budget provided by the company to actually focus on mitigating the negative environmental impact of AI, which limited their options.

\subsection{Impact of EU Regulations on AI Sustainability (RQ3)}
With the last theme, we studied how the related EU regulations, namely the AI Act and the CSRD, influence companies decision-making around AI sustainability.
Since the CSRD is based on company size, it did not apply to interviewees from companies smaller than 500 employees.
Table~\ref{table:regulations-overview} provides an overview of all created labels.

\begin{table}[ht]
\renewcommand{\arraystretch}{1.2}
    \centering
	\caption{Influence of Regulations: Subthemes and Labels (RQ3)}
	\label{table:regulations-overview}
	\begin{tabular}{lp{5.6cm}}
		Subtheme & Label\\
            \hline
            \hline
            \multirow{2}{*}{Impact of EU AI Act} & None [P1, P2, P3, P4, P5, P6, P8, P9, P10, P11] \\
                                                         & Considering [P7] \\
            \hline
            \multirow{3}{*}{Impact of CSRD}      & Complied [P3, P4, P6, P9, P10, P11] \\
                                                         & Not Applicable [P2, P5, P7, P8] \\
                                                         & None [P1] \\
            \hline
	\end{tabular}
\end{table}

\textbf{Impact of EU AI Act:}
Regarding the AI Act, 10 participants were either unfamiliar with it or had not taken any specific measures to comply with it (P1, P2, P3, P4, P5, P6, P8, P9, P10, P11).
All these participants expressed how the AI Act did not (yet) have any noticeable effect on how AI was used in their companies.
Product Owner P5 mentioned that, once the law would become effective, there would probably be more attention to it: \enquote{I think there is quite a lot of talk, especially now recently about the AI Act. There is a lot of interest for that. So, I think when it has been launched, then a lot more will be done to actually really look into it.}
P5 also highlighted that this regulation usually was more important regarding its risk categories, and for what applications AI can and cannot be used: \enquote{I think in general and especially when the AI Act was just announced, there is a lot of attention for these risk categories. I think that overshadowed a lot of the other topics in there.}
In similar fashion, P3 explained why they were still relatively complacent: \enquote{So, if you look at both the CSRD and the AI Act, it is something that we have to comply with as well. So, we have to consider that if we develop AI for our clients, it should be in line with the EU AI Act. Although this year, this calendar year, it is still allowed to ignore the Act, and then you have to do a lot of repairs just before the end of the year when the first part starts. So we should actually already take it into account preventively.}

While Co-Founder P7 mentioned how the AI Act was always in the back of their mind when working with AI models, P7 was not aware of the parts related to environmental sustainability: \enquote{The EU AI Act is important to us because it also puts restrictions on where you can and cannot use AI. Especially if you have GDPR-related data, you get classified to some extent. To be honest, the requirements related to the carbon emission, I was not aware of that.}
Similarly, P9 was also not aware of the environmental transparency that the EU AI Act proposes: \enquote{So, for the AI Act, like I just said, to be determined because I have not seen the environmental driver of the AI Act much so far. I have to get more involved there or acquainted with it, I think. It would be good if AI Act really has guidance on this, or ambitions.}

\textbf{Impact of CSRD:}
Based on their company size, seven participants needed to comply with the CSRD, namely P1, P3, P4, P6, P9, P10, and P11.
P1 was the only participant who did not know whether the company complied with the CSRD or not.
The other participants all explained that they did.
Nonetheless, Business Architecture Manager P6 was not yet satisfied with the content of their reporting: \enquote{So, at this moment, as far as I know, we are not obligatory reporting on the impact we are making with a project. But there is also intent to do that. We are doing a deal where we are staffing 20 people at one of our clients. It might make sense to look into, for example, how often our consultants are going to be at the client location. What kind of commutes are they doing? But also, they have 40 hours per week, their laptop running for this particular client. So, coming up with a report that provides some level of details in terms of, e.g., carbon footprint, but maybe also on how often they need to change their laptops and what kind of resources are impacted by that, whatever. We could definitely report on all of that, but at this moment, we do not.}
P9 was satisfied with the way the CSRD would be a step in the right direction for sustainable IT and said: \enquote{And in our own operations, IT is the dominant material. And so, yeah, we have to make transparent what is happening there. And reducing, well, this was already the whole idea of our sustainable IT department. So, basically, the CSRD is underpinning what we do. And that also enforces our role in the organization. So, we are quite happy with the CSRD. So, it's quite important, I would say.}

\section{Discussion}
In this section, we position our findings in relation to existing literature and discuss their interpretation and implications.

Regarding \textbf{RQ1}, the first theme revealed that the primary driver for adopting AI was efficiency.
Other mentioned drivers were security, personal passion, and the competitive market surrounding AI.
This is in line with the findings of \citet{radhakrishnan2020determinants}, where the facilitating factors of AI adoption were performance efficacy and efficiency, intrinsic motivation, and customer needs.
Once AI is adopted in the workplace, it is used for various tasks, such as data processing or as a chatbot.
AI has a significant contribution to many people’s work lives, so it is crucial to address its downsides to maximize its benefits.
During the adoption of AI, there was an awareness of the negative environmental impact of these AI tools, particularly of its carbon footprint and energy consumption.
It was only noted once that AI models also contribute to global water shortages~\cite{george2023environmental}, which supports previous findings on the narrow focus on carbon reduction in climate change discussions~\cite{savasta2014dangers}.

However, despite this awareness, sustainability concerns played no significant role in AI adoption decisions and were largely overshadowed by other drivers.
This indicates that companies do not (yet) perceive it as advantageous or important to consider environmental sustainability as a primary concern when deciding to adopt AI or not.
One important implication of this is that trying to convince companies to refrain from using AI for sustainability reasons will likely not be an effective strategy.
Instead, efforts should be spent on mitigating the negative environmental impacts of the used AI-based software.
Emphasizing the development of environmentally sustainable AI is therefore crucial, e.g., by raising industry awareness and integrating the principles of green AI into educational curricula.

Regarding \textbf{RQ2}, we highlighted that barely any interviewees monitored the negative environmental impacts of their AI tools.
Some participants used tokens as a measure of cost efficiency, which could indirectly be used to estimate the environmental impact.
However, this was not practiced by these companies.
This gap suggests a lack of perceived necessity for such monitoring.
Another possible reason may be the lack of easily accessible monitoring tools for environmental sustainability, i.e., we need to reduce barriers to using such tools.
Monitoring should be a low-hanging fruit as a first step on the path to more environmentally friendly AI.
The main obstacle mentioned by interviewees using off-the-shelf AI tooling was that AI service providers are not transparent enough about the negative environmental impact of their software products.
AI cloud providers really should make it easy for their clients to understand the energy use and carbon emissions of their software products and services.

The mitigation efforts we observed included prompt engineering, using smaller AI models, or reusing existing AI solutions.
While these efforts indicate a growing environmental awareness of AI use, this small number of example techniques stays far below available actionable Green AI knowledge and tools (see Section~\ref{sec:background}), i.e., there is significant room for improvement.
In addition to this knowledge gap, participants also indicated that sustainability efforts are largely postponed until after full AI integration has been achieved.
Sustainability was often only considered as an afterthought that was overshadowed by other concerns.
We thus recommend integrating sustainability considerations and mitigations as major parts of the AI adoption process.
 
Lastly, regarding \textbf{RQ3}, our results indicated that Green AI practices have not (yet) been significantly impacted by new regulations.
For example, our interviewees were not very familiar with the EU AI Act and very few have taken concrete actions in response to its environmental sustainability requirements.
Part of this can probably be explained by compliance not being fully mandatory yet, with practitioners being content to deal with it later.
However, another explanation might be that practitioners do not see these regulations as forceful enough regarding environmental sustainability, especially the AI Act.
In contrast, the CSRD seems to have a more tangible impact on larger companies, pushing them at least towards greater transparency, if not towards more sustainable AI practices.
While these regulations are positive steps in the right direction, they seem far from being very effective in the uptake of Green AI.

An intriguing observation emerged regarding the participants with less than a year of work experience.
These interviewees appeared to approach the study with a notable level of honesty, offering responses that felt genuine and unfiltered.
Unlike the more experienced participants, who often took more time to think before answering, the newer employees did not exhibit such hesitation.
This behavior of less experienced participants may suggest that they were less concerned with aligning their answers to any perceived expectations or normative standards.
The authenticity displayed by them is particularly valuable, as it provides clear insights into newly emerging work ethics, which might be hidden by the potential inclination to conform with usual practices in those with more experience.

In conclusion, this research highlights the critical role of accessible and actionable Green AI practices, and regulatory frameworks in addressing the environmental impact of AI.
While companies are beginning to recognize the importance of sustainable AI practices, significant changes are needed to integrate these considerations into their actual operational frameworks.
Our findings open new insights for further research and innovation, offering a richer understanding of the interplay between AI adoption, environmental sustainability, and regulatory influence.

\section{Limitations}
This study faced several limitations that should be acknowledged.
The first limitation is the sample size.
With 11 participants from 10 different companies of various domains, we had a decent amount of diversity, but still a limited number of data points.
While the responses to the interview questions were fairly consistent, the generalizability of our findings may be limited.
Moreover, we faced difficulties in accessing high-level executives and board members from large companies.
Their insights would have been invaluable, especially regarding strategic decisions and the adoption of AI at the highest levels of organizations.
Additionally, the majority of participants were based in the Netherlands, which may not fully represent the broader European or international context.

Another limitation is that our study was conducted at a single point in time.
Especially regarding the influence of regulations, a longitudinal study might provide much more reliable and detailed results.
This limited our ability to observe the long-term impacts of the EU AI Act.

Finally, the popularity or desirability of AI may have influenced the participants' responses, particularly concerning sustainability awareness and efforts.
Some interviewees, especially those in leadership positions, may have framed their company’s approach to AI and environmental sustainability in a more favorable light, consciously or unconsciously aligning their answers with perceived best practices.
This concurs with research indicating that professionals often tailor their responses to align with expected corporate narratives~\cite{paulhus2002use}.
Future studies could mitigate this limitation by incorporating observational data or case studies alongside self-reported interviews.
Nonetheless, the small number of applied Green AI practices was explicitly recognized by the participants, suggesting that they answered fairly truthfully.

\section{Conclusion}
As the power of AI continues to grow, it is important to further study how to profit from this technology without harming the environment and society.
However, it has been unclear which role environmental sustainability plays in the AI adoption decisions of practitioners, which Green AI practices are applied, and how regulations influence both of these.
Our study provides initial answers to these questions and highlights several areas for future research, e.g., developing mitigation strategies rather than slowing AI adoption.

\textbf{Broader sample.} Future research should include more industries and companies to help generalize our findings.
Ultimately, comparative studies across different regions could offer a broader perspective on the global Green AI perception and efforts.

\textbf{Long-term effects of regulations.} Examining the long-term effects of the EU AI Act and CSRD on AI practices and sustainability could provide valuable insights for policymakers.
Furthermore, research should explore the influence of corporate culture in fostering sustainable AI adoption.
Compliance should be driven by genuine commitment rather than regulatory avoidance, which requires a greater awareness of AI’s environmental impact.
Studies on attitudes and awareness can help identify key areas for improvements.
As AI continues to evolve, future research should assess the environmental implications of emerging AI models, hardware advancements, and software optimizations to inform best practices and sustainability guidelines.

\textbf{Improving monitoring and measurement techniques.} For greater awareness, it is important to keep applying and improving existing monitoring and measurement techniques, which should be accessible to both larger companies and individuals using AI.
Reliable and easy-to-use monitoring tools will encourage more effective mitigation efforts.

\textbf{Interdisciplinary collaboration.} Addressing the complex relationship between AI and environmental sustainability requires interdisciplinary collaboration.
Future research should involve experts from management science, computer science, software engineering, machine learning, environmental science, policy, social sciences, and business to develop comprehensive approaches to sustainable AI adoption.
These interdisciplinary perspectives can lead to more effective solutions and innovative strategies.

\section*{Acknowledgment}
We kindly thank all interviewees for their valuable time.

\bibliographystyle{IEEEtranN}
\bibliography{references}

\end{document}